\documentclass[11pt,a4paper]{article}
\usepackage{graphicx}
\usepackage{bm}
\newcommand{\be}{\begin{equation}}
\newcommand{\ee}{\end{equation}}
\newcommand{\beq}{\begin{eqnarray}}
\newcommand{\eeq}{\end{eqnarray}}
\newcommand{\beqs}{\begin{eqnarray*}}
\newcommand{\eeqs}{\end{eqnarray*}}
\newcommand{\lp}{\left( }
\newcommand{\rp}{\right) }
\newcommand{\llq}{\lq\lq}
\newcommand{\rrq}{\rq\rq}
\newcommand{\lb}{\left[ }
\newcommand{\rb}{\right] }

\begin{document}
\noindent {\Large \bf Kerr-Newman Solution as a Dirac Particle} \vskip 1.0cm
\noindent {\bf H. I. Arcos\footnote{Instituto de F\'{\i}sica Te\'orica,
Universidade Estadual Paulista, Rua Pamplona 145, 01405-900 S\~ao Paulo,
Brazil.}$^,$\footnote{Permanent address: Universidad Tecnol\'ogica de Pereira,
A.A. 97, La Julita, Pereira, Colombia.} and J. G. Pereira$^1$}

\vskip 2.0cm

\begin{abstract}
\noindent For $m^2 < a^2 + q^2$, with $m$, $a$, and $q$ respectively the source
mass, angular momentum per unit mass, and electric charge, the Kerr-Newman (KN)
solution of Einstein's equation reduces to a naked singularity of circular shape,
enclosing a disk across which the metric components fail to be smooth. By
considering the Hawking and Ellis extended interpretation of the KN spacetime, it
is shown that, similarly to the electron-positron system, this solution presents
four inequivalent classical states. Making use of Wheeler's idea of {\em charge
without charge}, the topological structure of the extended KN spatial section is
found to be highly non-trivial, leading thus to the existence of gravitational
states with half-integral angular momentum. This property is corroborated by the
fact that, under a rotation of the space  coordinates, those inequivalent states
transform into themselves only after a $4 \pi$ rotation. As a consequence, it
becomes possible to naturally represent them in a Lorentz spinor basis. The state
vector representing the whole KN solution is then constructed, and its evolution
is shown to be governed by the Dirac equation. The KN solution can thus be
consistently interpreted as a model for the electron-positron system, in which the
concepts of mass, charge and spin become connected with the spacetime geometry.
Some phenomenological consequences of the model are explored.
\end{abstract}
\vskip 1.0cm

\newpage
\section{Introduction}

The stationary axially-symmetric Kerr-Newman (KN) solution of Einstein's equations
was found by performing a complex transformation on the tetrad field for the
charged Schwarzschild (Reissner--Nordstr\"om) solution \cite{kerr,newman,newman2}.
For $m^2\geq a^2+q^2$, it represents a black hole with mass $m$, angular momentum
per unit mass $a$ and charge $q$ (we use units in which $\hbar = c = 1$). In the
so called Boyer--Lindquist coordinates $r,\theta,\phi$, the KN solution is given
by \cite{hellis} \be \label{metric}
 ds^2=dt^2-\frac{\rho^2}{\Delta} \, dr^2-(r^2+a^2) \sin^2\theta \, d\phi^2 -
\rho^2 \, d\theta^2-\frac{R r}{\rho^2} \, (dt - a\sin^2\theta \, d\phi)^2, \ee
where $\rho^2=r^2+a^2\cos^2\theta$, $\Delta=r^2-R r+a^2$ and $R=2m-q^2/r$. This
metric is invariant under the simultaneous changes $(t,a)\rightarrow(- t,-a)$,
$(m,r)\rightarrow(-m,-r)$, and separately under $q\rightarrow -q$. This black hole
is believed to be the final stage of a very general stellar collapse, where the
star is rotating and its net charge is different from zero.

The structure of the KN solution changes deeply when $m^2<a^2+q^2$. Due to the
absence of an horizon, it does not represent a black hole, but a circular naked
singularity in spacetime. This solution is of particular interest because it
describes a massive charged object with spin, and with a gyromagnetic ratio equal
to that of the electron \cite{newman}. As a consequence, several attempts to model
the electron by the KN solution have been made \cite{lopez,otro,israel}. In all
these models, however, the circular singularity is somehow surrounded by a massive
ellipsoidal shell (bubble), so that it is unreachable. In other words, the
singularity is considered to be non-physical in  the sense that the presence of
the massive bubble precludes its formation.

Inspired in the works by Barut \cite{barut1,barut2}, who tried to explain the {\it
Zitterbewegung} that appears in the electron's Dirac theory, and also in the works
by Wheeler \cite{mtw} about ``matter without matter'' and ``charge without
charge'', we will propose here that the KN solution, without any matter
surrounding it, can be consistently interpreted as a realistic model for the
electron. Our construction will proceed according to the following scheme. First,
by applying to the Kerr-Newman case the Hawking and Ellis extended  spacetime
interpretation of the Kerr solution \cite{hellis}, some properties of  the
classical KN solution for $m^2<a^2+q^2$ are reviewed. It is found that,  similarly
to the electron-positron system, the KN solution presents four  inequivalent
classical states. Then, an analysis of the topological properties of  the space
section of the extended KN spacetime is made. As already demonstrated in  the
literature \cite{fs}, the possible topologies of three-manifolds fall into two
classes, those which allow only vector states with integral spin, and those which
give rise to vector states having both integral and half-integral spins. In other
words, for a certain class of three-manifolds, the angular momentum of an
asymptotically flat gravitational field can present half-integral values,
revealing in this way the presence of a spinorial structure. As we are going to
see, this is exactly the case of the space section of the Hawking and Ellis
extended KN solution, when Wheeler's idea of charge without charge is  taken into
account. This will also become evident from the fact that each one of the
inequivalent KN states is seen to transform into itself only under a $4 \pi$
rotation, a typical property of  spinor fields. As a consequence of this property,
those states can naturally be represented in a Lorentz spinor basis. By
introducing  such a basis, the vector state representing the whole KN solution in
a rest frame  is obtained. Then, the general representation of this vector state
with a  nonvanishing momentum ($\vec p\neq 0$) is found, and its evolution is
shown to be  governed by the Dirac equation. It is important to remark that the
Dirac equation  will not be obtained in a KN spacetime. Instead, the KN solution
itself (for $m^2<a^2+q^2$) will appear as an element of a vector space, which is a
solution of the Dirac equation. Furthermore, as is usually done in particle
physics, the gravitational field produced by the electron---here represented by
the KN solution---will be supposed to be fast enough asymptotically flat, so that
the Dirac equation can be written in a Minkowski background spacetime. The above
results suggest that the KN solution can be consistently interpreted as a model
for the electron. At the final part of the paper, an analysis of some
phenomenological consequences of the model is presented, and a discussion of the
results obtained is made.

\section{The Extended KN Solution}

\subsection{Basic properties}

The KN solution for $m^2<a^2+q^2$ exhibits a true circular singularity of radius
$a$, enclosing a disk across which the metric components fail to be smooth. If the
center of the circle is placed at the origin of a Cartesian coordinate system, the
circular singularity coincides with the $xy$ plane, and the axial symmetry of the
solution (around the $z$-axis) becomes explicit. It should be remarked that, when
dealing with such solution, the concepts of mass and charge must be carefully used
because the presence of the singularity forbids one to apply both physical
concepts and laws along points in the border of the disk. The lack of smoothness
of the metric components across the enclosed disk can be remedied by considering
the extended spacetime interpretation of Hawking and Ellis \cite{hellis}, although
the circular singularity cannot be removed by this process. The basic idea of this
extension is to consider that our spacetime is joined to another one by the
singular disk. In other words, the disk surface (with the upper points considered
different from the lower ones) is interpreted as a shared border between our
spacetime, denoted by {\bf M}, and another similar one, denoted by {\bf M'}.
According to this construction, the KN metric components are no longer singular
across the disk, making it possible to smoothly join the two spacetimes, giving
rise to a single $4$-dimensional spacetime ${\mathcal M}$.

In what follows, we will assume the spacetime extension of Hawking and Ellis. This
means that the space sections of the spaces {\bf M} and {\bf M'} are joined
through the disks enclosed by the singularity. This linking can be seen as solid
cylinders going from one 3-manifold to the other (see Fig.~\ref{graf1}). The main
consequences of this interpretation are:
\begin{itemize}

\item{We can associate the electric charge of the KN solution on each $3$-manifold
with the net flux of a topologically trapped electric field, which goes from one
space to the other, as proposed by Wheeler \cite{mtw}. Although the electric field
lines seem to end at the singular ring (seen from either {\bf M} or {\bf M'}), the
equality of the electric charge on both sides of $\mathcal M$ tells us that no
electric field lines can ``disappear'' when going from {\bf M} to {\bf M'}, or
vice-versa. Then, in analogy with the geometry of the wormhole solution, there
must exist a continuous path for each electric field line going from one space to
the other. Furthermore, the equality of magnetic moment on both sides of $\mathcal
M$ implies that the magnetic field lines must also be  continuous when passing
through the disk enclosed by the singularity.}

\item{We can associate the mass of the solution with the degree of non-flatness of
the KN solution. Actually, the mass can be defined as the integral (we use in this
part the abstract index notation of Wald \cite{wald}) \be m = - \frac{1}{8\pi}
\int_{\sigma} \epsilon_{abcd} \, \nabla^c\xi^d = - \frac{3}{8\pi} \int_{V}
\nabla_{[e} \left\{\epsilon_{ab]cd} \nabla^c\xi^d\right\}, \ee where
$\epsilon_{abcd}$ is the spacetime volume element, $\xi^b =
\left(\partial/\partial t\right)^b$ is a timelike Killing vector-field, $V$ is a
spacelike hypersurface, and $\sigma$ is the 2-sphere boundary of $V$. From this
expression we obtain \be m =\frac{1}{4\pi} \int_{V} R_{ab} \, n^a \, \xi^b \, dV,
\label{extra3} \ee where $n^a$ is a unit future-pointing vector normal to $V$, and
$dV$ is the differential volume element which, in terms of the Boyer-Lindquist
coordinates, reads \be \label{difvol} dV = \frac{1}{2} \left(a^2 + 2 r^2 + a^2
\cos2\theta \right) \sin\theta \; dr\; d\theta\; d\phi. \ee Equation
(\ref{extra3}) shows that $m$ depends on the Ricci curvature tensor of spacetime.
This equation was obtained by Komar \cite{komar} and it is valid for all
stationary asymptotically flat spacetimes. It is important to remark that the
volume of integration must be taken either with $r>0$ or with $r<0$. Furthermore,
we can see from Eq.~(\ref{difvol}) that in the {\bf M} side ($r$ positive), the
mass $m$ is positive, whereas it is negative in the {\bf M'} side ($r$ negative).
Notice that the signs of $\xi^b$ and $n^a$ are not fixed in $\mathcal M$ since
both of them can be either positive or negative. It should also be noticed that
the mass $m$ is the {\em total} mass of the system, that is, the mass-energy
contributed by the gravitational and the electromagnetic fields are already
included in $m$ \cite{ohanian}.}
\end{itemize}

Now, using for $a$, $m$ and $q$ the experimentally known electron values, we can
write the total internal angular momentum $L$ of the KN solution on either side of
$\mathcal M$ as \be \label{cons3} L = m \, a, \ee which for a spin $1/2$ particle
assumes the value $L = 1/2$. It is then easy to see that the disk has a diameter
equal to the Compton wavelength $\lp\lambda/ 2\pi\rp_e = 1/m$ of the electron, and
consequently the angular velocity $\omega$ of a point in the singular ring turns
out to be \be \omega = 2 \, m, \ee which corresponds to Barut's {\it
Zitterbewegung} frequency \cite{barut1} for a point-like electron orbiting a ring
of diameter equal to $\lambda_e$. Therefore, if one takes the KN solution as a
realistic model for the electron, it shows from the very beginning a classical
origin for mass, electric charge and spin magnitude, as well as a gyromagnetic
ratio $g=2$. It should be remarked that, differently from previous models
\cite{lopez,otro,israel}, we are not going to suppose any mass-distribution around
the disk, nor around its border. Instead, we are going to consider a {\em pure}
(empty) KN solution where the values of mass, charge and spin are directly
connected to the space topology.

\subsection{Topology of the KN extended spacetime}

\begin{figure}[h]
\begin{center}
\includegraphics[height=6cm,width=9cm]{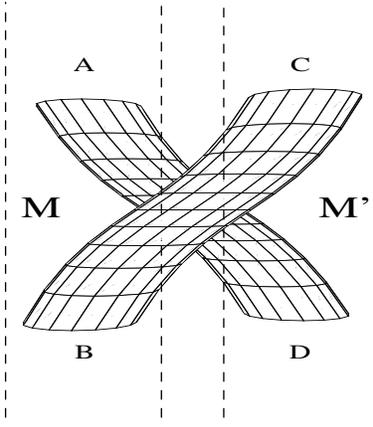}
\end{center}
\caption{To better visualize the intrinsic geometry of the 3-dimensional KN
manifold, the KN disk is drawn as if it presented a finite thickness, and
consequently there is a space separation between the upper and lower surfaces of
the disk. The left-hand side of each configuration state represents the upper and
lower surfaces of the disk in {\bf M}, whereas the right-hand side represents the
upper and lower surfaces of the disk in {\bf M'}. The lower B surface in {\bf M}
must be joined with the upper C surface in {\bf M'}, and the lower D surface in
{\bf M'} must be joined with the upper A surface in {\bf M}.} \label{graf1}
\end{figure}
\begin{figure}
\begin{center}
\includegraphics[height=8cm,width=8cm]{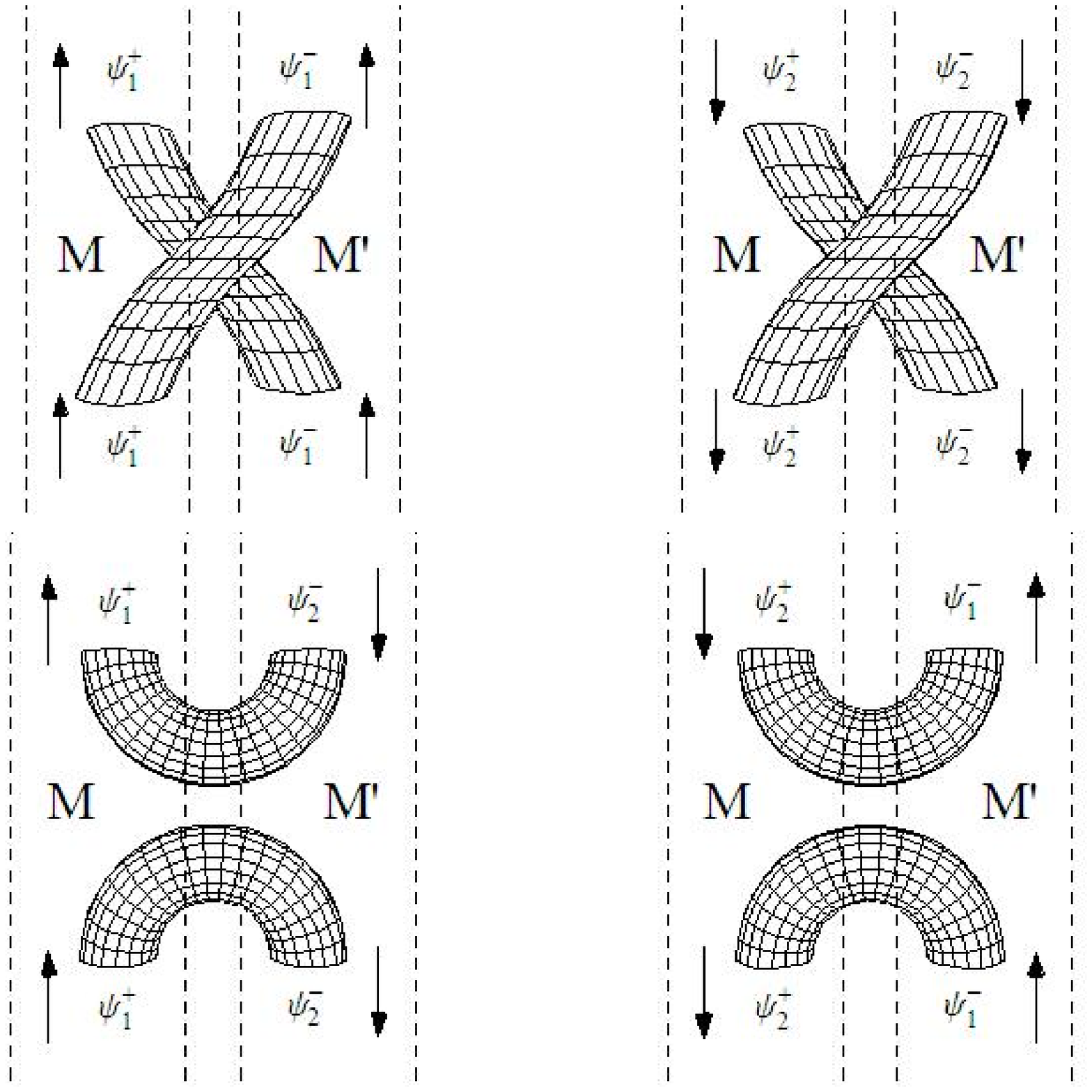}
\end{center}
\caption{The four possible geometric configurations of KN states for a specific
value of the electric charge. The KN disk on each space, that is, on {\bf M} and
on {\bf M'}, is placed on the $z=0$ plane. The arrows indicate the sense of the
spin vector.} \label{graf2}
\end{figure}
By a simple analysis of the structure of the extended KN metric, it is possible to
isolate {\em four} physically inequivalent states on each side of ${\mathcal M}$,
that is, on {\bf M} and {\bf M'}. These states can be labeled by the sense of
rotation ($a$ can be positive or negative), and by the sign of the electric charge
(positive or negative). Before a spin rotation axis is chosen, these states are
equivalent (up to a rotation) but after choosing it they are physically different.
If we place the KN disk in the $xy$ plane of a Cartesian coordinate system, the
spin vector will be either in the $+z$ or in the $-z$ direction. Now, each one of
these inequivalent solutions in {\bf M} must be joined continuously through the KN
disk to another one in {\bf M'}, but with opposite charge. Since we want a
continuous joining of the metric components, this matching must take into account
the sense of rotation of the rings. Through a spatial separation between the upper
and lower disks on each manifold, these joinings can be drawn as solid cylinders
(see Fig.~\ref{graf1}), which makes explicit the difference between both disks.
For the sake of simplicity, we are going to consider only one of the two possible
electric charges on {\bf M} ($q<0$, for example)\footnote{Two signs for the
electric charge $q$ in {\bf M} or {\bf M'} are allowed since the KN metric depends
quadratically on $q$.}. In Fig.~\ref{graf2}, the tubular joinings between {\bf M}
and {\bf M'}, just as in Fig.~\ref{graf1}, are drawn, but now taking into account
the different spin directions in each disk, which are drawn as small arrows. The
only differences between these configurations are the orientation of the spin
vector and the geometry of the tubes.

Now, in order to fully understand the topology of the spatial section of the KN
spacetime, let us obtain its spatial metric, that is, the metric of its 3-
dimensional space section. As the KN metric has non-zero off-diagonal terms, the
correct form of the 3-dimensional infinitesimal interval is \cite{landau}
\be\label{3metric} d\gamma^2=\lp -g_{ij}+\frac{g_{oi}g_{oj}}{g_{oo}}\rp dx^i\,dx^j
\equiv \gamma_{ij} \, dx^i\,dx^j, \ee where $i,j=1,2,3$. Applying this formula to
the KN metric (\ref{metric}), we obtain \beq\nonumber
d\gamma^2&=&\lp\frac{r^2+a^2\cos^2\theta}{r^2-2mr+a^2+q^2}\rp dr^2 + \lp
r^2+a^2\cos^2\theta\rp d\theta^2 \\ \label{3metrickn} &+&
\lp\frac{(r^2+a^2\cos^2\theta)\sin^2\theta}{r^2+a^2\cos^2\theta -2mr +q^2}\rp\lp
r^2+a^2-2mr+q^2 \rp d\phi^2. \eeq The first thing to notice is that the spatial
components are finite in the ring points, that is, $r=0$ and $\theta=\pi/2$. This
does not mean that the singularity is  absent. Rather, it means that only the
metric derivatives are singular, not the metric itself. We can thus conclude that
the spatial section of the KN solution has a well defined topology. In fact, the
distance function
\[
d(p,q) = \int_p^q\sqrt{\gamma(u)} \; du
\]
is easily seen to be finite for any nearby points $p$ and $q$ of the space. The
basic conclusion is that the KN space section, or ${\mathcal M}^3$, has a well
defined topological structure, and is consequently a topological space.

\begin{figure}
\begin{center}
\includegraphics[height=3.8cm,width=15cm]{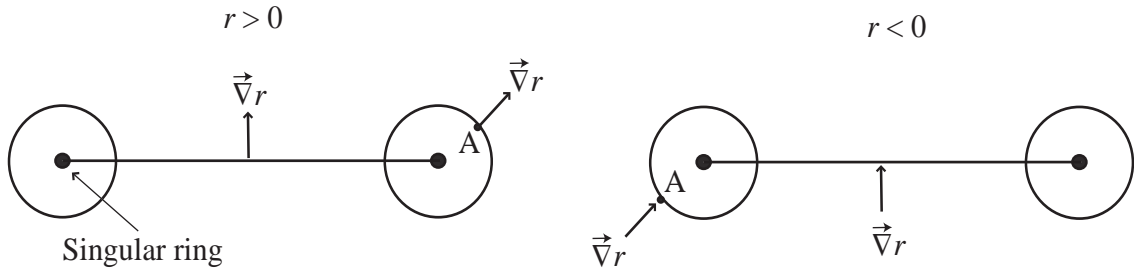}
\end{center}
\caption{Tubular-like regions around the singular ring, which is to be excised.
Several $\vec\nabla r$ directions are also depicted, which show how the borders in
the positive and negative $r$ sides can be continuously glued.} \label{graf3b}
\end{figure}

In spite of presenting a well defined topological structure, the space ${\mathcal
M}^3$ is not locally Euclidean everywhere. To see that, let us calculate the
spatial length ${\mathcal L}$ of the border of the disk $r=0$. It is given by
\be\label{lofb} {\mathcal L} = \int_0^{2\pi}d\gamma, \ee where the integral is
evaluated at $r=0$ and $\theta=\pi/2$. As a simple calculation shows, it is found
to be zero, which means that the border of the disk is topologically a single
point of ${\mathcal M}^3$. Therefore, an open ball centered at the point $r=0$,
$\theta=\pi/2$ will not be diffeomorphic to an open Euclidean ball, and
consequently the space ${\mathcal M}^3$ will not be locally Euclidean on the
border of the disk $r=0$. This problem can be solved by using Wheeler's concept of
the electric charge. According to his proposal, the electric field lines never end
at a point: they are always continuous. It is the non-trivial topology of
spacetime which, by trapping the electric field, mimics the existence of charge
sources. Applying this idea to the KN solution, we see that the electric field
lines end at the singular ring only from the 4-dimensional point of view. From the
3-dimensional point of view, they end at a point (the border of the disk). Now, if
this point is not the end of the electric field lines, then they must follow a
path to the side with $r$ negative. This situation is quite similar to that of the
Reissner-Nordstr\"om solution, where the electric field lines can be continued to
the negative $r$ side by writing the solution in Kruskal-Szekeres coordinates
\cite{mtw} (it should be noted, however, that in this case the solution is not wholly
static since the timelike coordinate changes to spacelike at small distances from
the singularity). In the same way as in the Reissner-Nordstr\"om solution, we can
excise a neighborhood of the point $r=0$ of the KN solution, and join again the
resulting borders. In the KN case, considering the values of the electron mass,
charge and spin, the time coordinate keeps its timelike nature at all points of the
space, so the solution remains stationary after the excision procedure.
Furthermore, from the causal analysis of the solution we  can choose to excise
precisely the torus-like region around the singularity where there exist non-causal
closed timelike curves. From the metric (\ref{metric}), it is is easy to see that
the coordinate $\phi$ becomes timelike in the region where the following inequality
is fulfilled
\cite{carter}:
\be\label{ineqc}
r^2+a^2-\lp\frac{q^2-2mr}{r^2+a^2\cos^2\theta}\rp a^2\sin^2\theta < 0.
\ee
By removing this region, the KN spacetime becomes causal. As already said, this
region has a simple form: it is tubular-like and surrounds the singular ring on
the negative and positive $r$ sides. When the values of $a$, $q$ and $m$ are
chosen to be those of the electron, the surface of the tubular-like region is
separated from the singular ring by a distance of the order of $10^{-34}$ cm. At
these infinitesimal distances, topology changes are predicted to exist, so it is not
unexpected to have changes in the connectedness of spacetime topology. Wheeler's
idea can then be implemented in the KN case. This means to excise the infinitesimal
region from the positive and negative $r$ sides, and then glue back the manifold
keeping the continuity of both electric field lines and metric components. A simple
drawing of the region to be excised can be seen in Fig.~\ref{graf3b}, where the
direction of the gradient of $r$ has been drawn at several points, and the region's
size has been greatly exaggerated.
As an example, note that the point $A$ on the positive $r$ side must be glued to
the point $A$ on the negative $r$ side. If we continue to glue all points of the
tubular border, we obtain a continuous path for the electric field lines that flow
from one side of the KN solution to the other side. Wheeler's idea is then fully
implemented, yielding a 3-dimensional spatial section ${\mathcal M}^3$ which is
everywhere locally Euclidean, and consequently a Riemannian manifold.

For the sake of completeness, we determine now the form of the surface obtained by
joining the points of the tubular borders in a metric-continuous way. To do it, we
must define the application ${\cal A}: S^1\rightarrow S^1$, which is
constructed in the following way: draw the two $S^1$ of Fig.~\ref{graf3b}, but now
centered at the points $(a,0,0)$ and $(-a,0,0)$ of a Cartesian coordinate system.
The application ${\cal A}(p\in S^1)$ is then defined as ${\cal A}(x,0,z)=(-x,0,-
z)$, where $(x,0,z)$ are the coordinates of $p$. The form of this application is
deduced from the restriction of joining the tubular borders in a metric-continuous
way. The surface determined by the joinings is then generated by the quotient
space of $S^1$ by the equivalence relation $p\sim {\cal A}(p)$, and by a rotation
of $\pi$ of the circles around $z$. This surface coincides with the well known
{\em Klein Bottle} (denoted by $U_2$), as can be easily verified in
Ref.~\cite{carmo}. It is important to remark that the Klein bottle has also been
found by Punsly \cite{punsly} in the context of the Kerr solution, through a
metric extension procedure which eliminates the singular ring. However, in our
case we have a physical justification for the excision procedure: the continuity
of the electric field lines.

\begin{figure}
\begin{center}
\includegraphics[height=8cm,width=5cm]{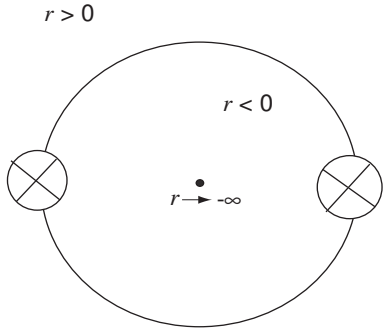}
\end{center}
\vskip -3.8cm
\caption{View of the total $2$-dimensional symmetric section of ${\mathcal M}^3$
obtained through the map $r \rightarrow e^r$. The big circle represents the
surface $r=0$. The ring singularity is located in the equator, and the small
circles represent the infinitesimal regions to be excised around the singularity.}
\label{grafextra}
\end{figure}

The elimination of the singular ring can be better understood by considering the
map $r \rightarrow e^r$, which makes possible to see both sides of the $r=0$
surface. This map is shown in Fig.~\ref{grafextra}, with the circles representing
the small excised neighborhoods around the ring singularity. The full picture of
the resulting $3$-manifold is obtained by rotating the plane of the figure by
$\pi$. It is clear from this figure that the resulting manifold will be multiple
connected since any path encircling the excised region cannot be contracted to a
point. The borders of the excised regions must be joined in such a way to make the
radial component of the metric continuous, as discussed in the analysis of
Fig.~\ref{graf3b}. The other components of the metric are equal on the excised
borders (by rotational symmetry), so that they can be always continuously matched.
According to the above construction, ${\mathcal M}^3$ is a non-trivial
differentiable three-manifold. This three-manifold can be seen as a connected sum
of the form \be \label{cs} {\mathcal M}^3 = \mathbf{R}^3 \sharp
\mathbf{K}^3\sharp\mathbf{R}^3, \ee where the first $\mathbf{R}^3$ represents the
asymptotic space section of {\bf M}, the second $\mathbf{R}^3$ represents the
asymptotic space section of {\bf M'}, and the 3-space $\mathbf{K}^3$ is the
manifold formed by the one-point compactification of ${\mathcal M}^3$, which is
obtained by adding the two points at $\{-\infty,\infty\}$. Then, each 2-sphere,
respectively at $\pm \infty$, must be taken as a point of $\mathbf{K}^3$. If the
connection between the disks were removed, $\mathbf{K}^3$ would become  homotopic
to two disjoint $3$-spheres. This comes from the fact that $\mathbf{R}^3 \bigcup
\{\infty\} \simeq S^3$. But, as far as the joinings are present, $\mathbf{K}^3$ is
not simply connected because there exist loops (those  surrounding the $U_2$
surface) not homotopic to the identity. In fact, the first  fundamental group of
$\mathbf{K}^3$ is found to be $\pi_1 (\mathbf{K}^3) = \mathbf{Z}$. Furthermore,
the second fundamental group of $\mathbf{K}^3$ is found to be $\pi_2(\mathbf{K}^3)
= \{e\}$.

\section{Existence of Half-Integral Angular Momentum States}
\subsection{Topological conditions}

In order to exhibit gravitational states with half-integral angular momentum, a
$3$-manifold ${\mathcal M}^3$ must fulfill certain topological conditions. These
conditions were stated by Friedman and Sorkin \cite{fs}, whose results were
obtained from a previous work by Hendriks \cite{hen} on the obstruction theory in
$3$ dimensions. In order to understand Hendrik's result, it is convenient to
divide the manifold ${\mathcal M}^3$ into an interior (${\mathcal M}^3_I$) and an
exterior (${\mathcal M}^3_E$) regions in such  a way that ${\mathcal M}^3_I \cap
{\mathcal M}^3_E$ is a spherical symmetric  shell. After that, one defines a
rotation by an angle $\alpha$ of the submanifold ${\mathcal M}^3_I$ with respect
to ${\mathcal M}^3_E$ as a three-geometry obtained  by cutting ${\mathcal M}^3_I$
at any sphere $S^2\subset {\mathcal M}^3_I \cap {\mathcal M}^3_E$, and re-
identifying (after rotating) the inner piece with the outer. Then, one looks for a
diffeomorphism that takes the final three-geometry, obtained after a rotation of
$\alpha=2 \pi$, to the initial one, characterized by $\alpha$ equal to $0$. If
this diffeomorphism can be deformed to the identity, the gravitational states
defined on the manifold can have only integral angular momentum. If the
diffeomorphism cannot be deformed to the identity, then half-integral angular
momentum gravitational states do exist. This diffeomorphism was called by Hendriks
a {\it rotation parallel to the sphere}, and it will be denoted by $\rho$.

Hendriks' results can then be summarized in the following form. If the division
into an exterior and interior region is not possible, then $\rho$ cannot be
deformed into the identity. Physically, this means that if {\bf M} and {\bf M'}
are joined not only at the surface $r=0$, but also at any other place, ${\mathcal
M}^3$ can exhibit half-integral angular momentum states, since in this case there
would not exist interior and exterior regions to a shell that encloses the $r=0$
surface. On the other hand, if the division is possible, then ${\mathcal M}^3$
will exhibit only {\em integral} angular momentum states only if it is a connected
sum of compact three-manifolds (without boundary),
\[
{\mathcal M}^3 = \mathbf{R}^3 \sharp M_1 \sharp \ldots \sharp M_k,
\]
each of which (a) is homotopic to $P^2\times S^1$ ($P^2$ is the real projective
two-sphere), or (b) is homotopic to an $S^2$ fiber bundle over $S^1$, or (c) has a
finite fundamental group $\pi_1(M_j)$ whose two-Sylow subgroup is cyclic. In order
to exhibit {\em half-integral} angular momentum states, therefore, the 3-manifold
${\mathcal M}^3$ must fail to fulfill either one of these three conditions.

According to the decomposition (\ref{cs}), ${\mathcal M}^3$ can be seen as the
connected sum of two $\mathbf{R}^3$ and $\mathbf{K}^3$. Now, as the original
analysis of Hendriks was made for compact topological manifolds without boundary,
we have to compactify $\mathbf{K}^3$ by adding two points at infinity. Besides
compact, the resulting 3-space turns out to be without boundary (see \ref{b}).
Accordingly, Hendriks results can be used, and we can say that the manifold
${\mathcal M}^3$  will admit half-integral angular momentum states only if
$\mathbf{K}^3$ fails to fulfill one of the above conditions (a) to (c). Condition
(a) is clearly violated because, as $\pi_1(\mathbf{K}^3) = \mathbf{Z}$, and as
\[
\pi_1(P^2\times S^1) = \pi_1(P^2)\oplus\pi_1(S^1) = \mathbf{Z}_2\oplus\mathbf{Z},
\]
$\mathbf{K}^3$ cannot be homotopic to $P^2\times S^1$. Condition (b) is more
subtle, but it is also violated. In fact, as is well known \cite{st}, the number
of inequivalent bundles of $S^2$ over $S^1$ is just two: A trivial and a non-
trivial one. Since the non-trivial bundle is always non-orientable, $\mathbf{K}^3$
cannot be homotopic to this space since it is orientable by construction. The
trivial bundle $T^3$, on the other hand, is formed by taking the direct product of
$S^2$ with $S^1$. We then have
\[
\pi_1(T^3)=\pi_1(S^2\times S^1)=\pi_1(S^2)\oplus\pi_1(S^1)=\{e\}\oplus\mathbf{Z},
\]
which is formally the same as $\pi_1(\mathbf{K}^3)$. However, the second homotopy
group of $T^3$ is given by
\[
\pi_2(T^3)=\pi_2(S^2\times S^1)=\pi_2(S^2)\oplus\pi_2(S^1)=\mathbf{Z}\oplus\{e\},
\]
and as $\pi_2(\mathbf{K}^3) = \{e\}$, then clearly
$\pi_2(\mathbf{K}^3)\neq\pi_2(T^3)$. This shows that $\mathbf{K}^3$ cannot be
homotopically deformed to a bundle $S^2$ over $S^1$. Finally, as discussed in the
last section, condition (c) is also violated because the fundamental group
$\pi_1(\mathbf{K}^3) = \mathbf{Z}$ is infinite. We can then conclude that the {\it
KN spacetime does admit gravitational states with half-integral angular momentum}.
More precisely, we can conclude that it admits gravitational states with spin 1/2.

\subsection{Behavior under rotations}

By using the definition of $\rho$ introduced earlier in this section, we can
proceed to analyze the effect of a rotation in the region around $r=0$ of the
manifold ${\mathcal M}^3$. The following analysis apply to anyone of the two
possible interpretations, {\bf M} joined to {\bf M'} through $r=0$ only, or
through various other points. One has to choose a spherical shell centered on any
one of the two sides of the $r=0$ surface. After choosing the shell one must look
at the effect of a rotation on the 3-geometry of the manifold. For simplicity, we
choose the positive side of de surface $r=0$ centered on a Cartesian coordinates
system and a shell centered on $(0,0,0)$ with a radius large enough so that the
geometry outside the shell can be taken as flat.

\begin{figure}
\begin{center}
\includegraphics[height=7cm,width=12cm]{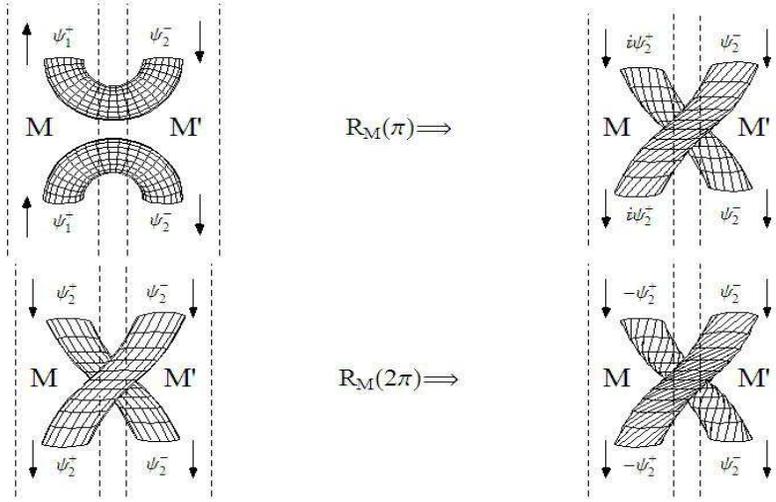}
\end{center}
\caption{The effect, on the KN states, of rotations around the $x$ axis of the
interior region of a shell enclosing $r=0$. The unwitting diffeomorphism can be
seen as a dilatation of the tubes that takes them  around one of the extreme
points of the cylinders.} \label{graf5}
\end{figure}

If we perform now a rotation by an angle $\alpha$ around anyone of the axis $x$,
$y$ or $z$ of the interior region of the shell, the effect on the 3-geometry is to
twist the cylindrical tubes of Fig.~\ref{graf1}. In the specific case of a
rotation around the $x$ axis, the twist of the tubes is shown in Fig.~\ref{graf5}
for $\alpha = \pi$ and $\alpha = 2 \pi$.
From this figure, it can be inferred that only after a $4\pi$ rotation the
twisting of the tubes can be undone by deforming and taking them around one of
their extreme points. In other words, only after a $4\pi$ rotation a
diffeomorphism connected to the identity does exist, which takes the metric of the
twisted tubes into the metric of the untwisted ones. We mention in passing that
the form of this diffeomorphism is equal to the one solving the well-known Dirac's
{\it scissors} problem \cite{pr}.

\begin{figure}
\begin{center}
\includegraphics[height=5cm,width=14cm]{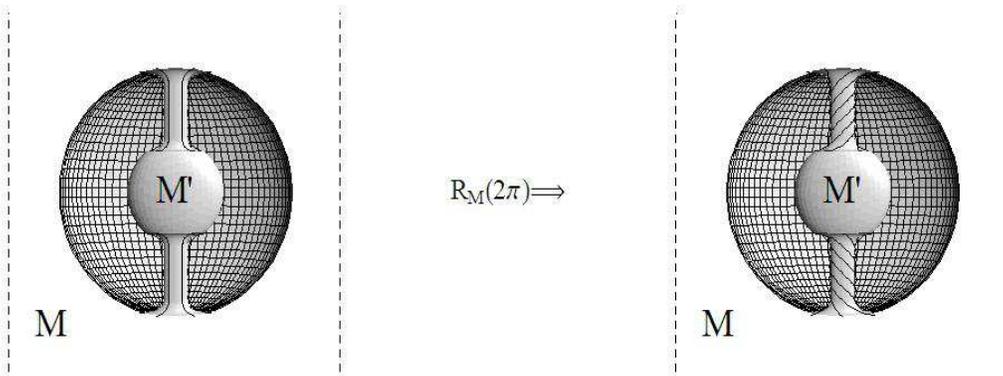}
\end{center}
\caption{By performing the coordinate transformation $r\rightarrow \ln R$, the KN
extended manifold can be represented as depicted in the figure. The central
surface represents the points with $R=1$, and {\bf M'} is represented by the inner
points. The outer surface represents also the points with $R=1$, but seen from the
{\bf M} side. The tubes joining the two spaces are drawn vertically. To the right,
it is represented the effect of a $2\pi$ rotation in the interior region of a
shell enclosing $r=0$.} \label{graf4}
\end{figure}

The effect of a rotation in the interior region of the chosen shell can also be
seen by performing first a transformation in the Boyer-Lindquist coordinates that
modifies the coordinate $r$ only: \be \label{res} (t,r,\theta,\phi)\rightarrow
(t,\ln R,\theta,\phi). \ee This transformation compactifies {\bf M'}, and takes
the points on the disk ($r=0$) into the points on the surface $R=1$. A simplified
form of the transformed three manifold ${\mathcal M}^3$ can be seen in
Fig.~\ref{graf4}. In this figure, the tubes joining the spaces {\bf M} and {\bf
M'} are drawn vertically. They must join the points of the inner surface (except
those points near the equator) with the points of the outer surface. The points of
{\bf M'} are those within the central surface, which is defined by $R=1$. It
should be remarked that the homotopy groups of ${\mathcal M}^3$ are not altered by
the transformation.

Now, a rotation in the interior region of a shell enclosing $r=0$ can be seen as a
twisting of the cylindrical tubes that connect the  two spaces. For a rotation by
an angle $\alpha = 2 \pi$, this twisting cannot be  undone by a diffeomorphism
homotopic to the identity, since the extreme points of  the cylinder should be
kept fixed for it to be connected to the identity. For a  rotation by an angle
$\alpha = 4\pi$, however, it is possible to untwist them because in this case
there exists a diffeomorphism homotopic to the identity that  untwist them, and at
the same time keeps the extreme points fixed. The form of  this unwitting
diffeomorphism can be found in page 309 of ref.\ \cite{mtw}. The fact that the
topological structure of the spatial section of the KN  manifold returns to its
initial state after a $4 \pi$ rotation is a more intuitive  way to see that this
space admits spinorial states.

\section{Algebraic Representation of the KN States}
\subsection{Spinor states}

Following Ref.~\cite{fs}, we denote by $\Upsilon({\mathcal M}^3)$ the space of
asymptotically flat positive-definite three-metrics $g_{ab}$ on ${\mathcal M}^3$.
Since the metric on ${\mathcal M}^3$ is fixed (up to a diffeomorphism), different
points in $\Upsilon({\mathcal M}^3)$ represent three-metrics which differ only by
the geometry of the joining between the sides of the $r=0$ surface. We define a
state vector $\psi$, in the Schr\"odinger picture, as a functional on the space
$\Upsilon({\mathcal M}^3)$. The generalized position  operator $\widehat g_{ab}$
is then defined as \be \label{gpo} \widehat g_{ab} \psi(g) = g_{ab} \psi(g), \ee
which means that, for every point of $\Upsilon({\mathcal M}^3)$, we have a
different state vector $\psi(g)$. Now, from  the discussion of the last section,
we can say that (under rotations) a path in $\Upsilon({\mathcal M}^3)$ is closed
if and only if the parameter of the path  varies from $0$ to $4\pi$. Furthermore,
since the points of $\Upsilon({\mathcal  M}^3)$ are in one to one correspondence
with the states $\psi(g)$, we find that  the effect of a $2\pi$ rotation on
$\psi(g)$ is not equal to the identity  operation:
\[
\widehat R(2\pi)\psi(g)=\psi(R(2\pi)g)=\psi(g'\neq g)\neq \psi(g).
\]

An adequate linear representation for the states $\psi(g)$ is one that carries, in
addition to the informations about mass and charge, also information on the non-
trivial behavior under rotations of the states representing the KN solution. As
the state $\psi(g)$ depends on the metric $g_{ab}$, it is not a simple task to
infer its general form. As a first step, we can separate a general $\psi(g)$ into
a part ($\psi^+$) defined on the positive $r$ coordinates, and another ($\psi^-$)
defined on the negative $r$  coordinates: \be \label{sep} \psi = a_+\psi^+ + a_-
\psi^-. \ee Furthermore, if we choose the spin direction along the $z$-axis, we
have two possibilities for it (see Fig.~\ref{graf2}). Therefore, we can write \beq
\label{sep2}
\psi^+&=&b_1\psi^+_1+b_2\psi^+_2 \\
\label{sep3} \psi^-&=&c_1\psi^-_1+c_2\psi^-_2. \eeq This superposition is {\it
necessary} because only after a measurement of the spin direction we know for sure
the sign of $a$ in the metric (\ref{metric}). Replacing (\ref{sep2}) and
(\ref{sep3}) into (\ref{sep}), we obtain \be\label{sepf} \psi=a_+\lp
b_1\psi^+_1+b_2\psi^+_2\rp + a_-\lp c_1\psi^-_1+c_2\psi^-_2\rp. \ee We want also
that the state $\psi(g)$ be an eigenvector of both the energy and the spin
operators. This means that \beq
S_z\psi^\pm_1&=&s_z\psi^\pm_1=\frac{1}{2}\psi^\pm_1 \\
S_z\psi^\pm_2&=&s_z\psi^\pm_2=-\frac{1}{2}\psi^\pm_2 \\
\label{oq3} H\psi^+&=&- i\partial_t\psi^+=m\psi^+\\ \label{oq4} H\psi^-&=&-
i\partial_t\psi^-=-m\psi^- \eeq where $S_z$ is the spin operator along the $z$
direction, $H$ is the energy operator, and $m$ is the mass of the KN solution. In
these relations we have implicitly used the correspondence between mass and energy
(remember that $m$ is negative on the negative $r$ sector of ${\mathcal M}^3$).

Now, as a consequence of the fact that an observer in the positive $r$ side of
$r=0$, as well as one in the negative $r$ side, sees a state vector that
transforms into itself only after a $4\pi$-rotation, we can naturally represent
these states in a spinor basis of the Lorentz group SL(2, $\mathbf{C}$). More
specifically, each one of the four inequivalent states defined in the positive $r$
side can be taken as Weyl spinors transforming under the $({1}/{2},0)$
representation, and those defined in the negative $r$ side as Weyl spinors
transforming under the $(0,{1}/{2})$ representation of the Lorentz group (see
\ref{a} for a more detailed discussion). Furthermore, according to
Eqs.~(\ref{oq3}) and (\ref{oq4}), the linear representation for $\psi$ must also
contain a part proportional to a complex exponential of energy multiplied by time.
Finally, it is important to notice that the representation cannot mix $\psi^+$ and
$\psi^-$ as they are defined on different spatial regions. With these provisos, we
are led to the following representation for $\psi(g)$: \be \label{iner1} \psi(g)=
a_+ \left[ b_1 e^{iEt}\lp\begin{array}{c} 1
\\ 0 \\0 \\ 0\end{array}\rp +b_2 e^{iEt}\lp\begin{array}{c} 0 \\ 1
\\0 \\ 0\end{array}\rp \right] + a_- \left[ c_1
e^{-iEt}\lp\begin{array}{c} 0 \\ 0 \\1 \\ 0\end{array}\rp+c_2 e^{-
iEt}\lp\begin{array}{c} 0
\\ 0 \\0 \\ 1\end{array}\rp \right].
\ee This can be considered as the most simple linear representation of a general
$\psi(g)$ associated with the KN solution (at rest).

\subsection{Evolution equation}

If we want the above solution to represent a particle, we need it to be an
eigenstate of momentum (or position). This is not easy because the \llq
position\rrq\/ of the \llq particle\rrq\/ is not defined by a simple point in
spacetime. So, for example, the usual momentum operator $-i\vec\nabla$ is of no
use because it is defined for point-like particles. To solve this problem, we need
to get an approximate representation for $\psi(g)$, valid in the limit of long
distances from the singularity. In this limit, the KN solution is supposed to
converge to the metric produced by a spinning structureless point particle. Only
in this case the momentum operator $-i\vec\nabla$ becomes well suited for defining
the momentum of the particle. In this limit, we can also consider the background
metric as flat, which means to consider $\eta_{\mu\nu} = {\rm diag}(-1,1,1,1)$ as
the spacetime metric. Using the form of $\psi(g)$ as given by Eq.\ (\ref{iner1}),
we find that an adequate dependence on the momentum of the particle is given by
the exponential $i\vec p\cdot\vec x$. This is due to the fact that, in this case,
the two exponential factors combine to give a covariant expression, and at the
same time the state becomes an eigenvector of momentum. The most general state
$\psi_p(g)$ is then given by \be \label{iner2}
 \psi_p(g)= a_+e^{-ip_\mu x^\mu}
\left[ b_1\lp\begin{array}{c} 1 \\ 0 \\0 \\ 0\end{array}\rp
+b_2\lp\begin{array}{c} 0 \\ 1 \\0 \\ 0\end{array} \rp \right] + a_-e^{ip_\mu
x^\mu} \left[ c_1\lp\begin{array}{c} 0 \\ 0 \\1 \\
0\end{array}\rp+c_2\lp\begin{array}{c} 0 \\ 0
\\0 \\
1\end{array}\rp \right]. \ee

Now, from Eqs.~(\ref{oq3}) and (\ref{oq4}), we can write the evolution equation
for the KN state as \be \label{dirac} \frac{1}{i}\partial_t \psi_p=H\psi_p \equiv
E\lp\begin{array}{cc} I_2 &0\\0&- I_2 \end{array}\rp\psi_p\equiv E \, \beta \,
\psi_p, \ee where $I_2$ is the $2\times 2$ identity matrix. The minus sign of the
lower components is a reflection of the fact that the lower components of the
vector state have negative energy. A more convenient form of the evolution
equation can be obtained by performing a unitary transformation. We write this
transformation, which is a particular case of the well known {\em Foldy-
Wouthuysen} transformation \cite{gross}, in the form \be \label{transf} \mathbf{U}
= \sqrt{\frac{E+m}{2 E}} \; \lp
\begin{array}{cc}
I_2 &-\frac{\sigma_ip_i}{E+m} \\
\frac{\sigma_ip_i}{E+m}& I_2 \end{array}\rp. \ee It can then be easily verified
that \be \Psi_p = \mathbf{U} \; \psi_p \ee is a solution of the modified evolution
equation \be \label{diracformal} \frac{1}{i}\partial_t \Psi_p=\mathbf{H} \, \Psi_p
\equiv \lp\alpha_i p_i+\beta  m\rp\Psi_p, \ee with \be \mathbf{H} = \mathbf{U} H
\mathbf{U}^\dag \ee the transformed Hamiltonian. As is well known, Eq.
(\ref{diracformal}) is the standard form of Dirac's relativistic equation for the
electron. The basic conclusion is that the KN solution of Einstein's equation can
be represented by a state vector that is a solution of the Dirac equation. Besides
exhibiting all properties of a solution of the Dirac equation, the KN state
provides an intuitive explanation for mass, spin and charge. Furthermore, it
clarifies the fact that, during an interaction, both positive and negative energy
states contribute to the solution of the Dirac equation. This means that it is not
possible to describe interacting states as purely positive or purely negative
energy states since, as the extended KN solution explicitly shows, the two energy
states are topologically linked. On the other hand, it is possible to describe a
free, moving, positive-energy (negative-energy) state without considering
negative-energy (positive-energy) components. It should be remarked that there
exists an arbitrariness in this terminology. In fact, what we call a negative-
energy state and a positive-energy state depends on which side of the KN solution
we are supposed to live in. Furthermore, as the electric charge enters
quadratically in the KN metric, it is not possible to say in which side of the KN
solution an ideal observer is, and consequently what we call positive or negative
energy state is also a matter of convention.

\section{Some Phenomenological Tests}

We look now for some experiments where the effects of the singularity would become
manifest. We begin by noticing that, for symmetry reasons, the electric dipole
moment of the KN solution vanishes identically, a result that is within the limits
of experimental data \cite{edm}. Another important point is that the uncertainty
principle precludes one to localize the electron in a region smaller than its
Compton wavelength without producing virtual pairs originated from the large
uncertainty in the energy. Since we are proposing an extended electron model with
the size of its Compton wavelength, the question then arises whether it
contradicts scattering experiments that gives a limit to the extendedness of the
electron as smaller than $10^{-18}$cm. This is a difficult question because this
model describes the electron as a nontrivial topological structure with a trapped
electromagnetic field. As a consequence, its interaction with other electrons must
be governed by the coupled Einstein--Maxwell equations. Even though, a simplified
answer to this question can be given by noticing that a boost (in the spin
direction) transforms the Kerr--Newman parameters $m$ and $a$ according to
\cite{buri2} \be \label{fff} a' = a\sqrt{1-v^2}; \qquad m'=\frac{m}{\sqrt{1-v^2}},
\ee where $v$ is the boost velocity. It should be clear that $a$ and $m$ are
thought of as parameters of the KN solution, which only asymptotically correspond
respectively to angular momentum per unit mass and mass. Near the singularity, $a$
represents the radius of the singular ring, which according to Carter is
unobservable \cite{carter}. The above ``renormalization'' of the KN parameters has
been discussed by many authors \cite{norma}, being necessary to maintain the
internal angular momentum constant. As a consequence, to a higher velocity, there
might correspond a smaller radius of the singular ring. With this renormalization,
it is a simple task to verify that, for the usual scattering energies, the
resulting radius is within the experimental limit for the extendedness of the
electron. According to these arguments, therefore, the electron extendedness will
not show up in high-energy scattering experiments. This extendedness will show up
only in low energy experiments, where the electrons move at low velocities.

Let us then look for a simple low energy test involving interactions with other
particles, or electromagnetic fields. Take, for example, a pair of electrons
confined to a two dimensional plane $D$. If a strong magnetic field perpendicular
to the plane is applied, the spin vectors of the electrons will align with the
magnetic field. This means that the KN disk will be coplanar with $D$. The
magnetic flux $\Phi$ through the plane is given by \be \label{flux}
\Phi=\int_D{\vec B \cdot d\vec s}=\int_D{\vec\nabla\times\vec A(z) \cdot d\vec
s}=\oint_{\partial D}A(z)dz, \ee where $\vec B$ is the magnetic field, $\vec A$ is
the vector potential defined on $D$, and $z=x+iy$ are complex coordinates for the
plane. The border of $D$ will have three parts: An external part, and the border
enclosing the KN disk for each electron. Taking periodic boundary conditions on
the external border,\footnote{In practice, the plane $D$ must be finite for the
electrons to be confined.} we are left with two disconnected borders only. In such
multiply-connected spacetime, the vector potential $\vec A(z)$ is not uniquely
defined since there exist two other closed one-forms $\zeta_k^{-1} d\zeta_k$,
$k=1,2$, with the property \cite{fqhe1} \be \label{onefor} \int_{\partial D_k}
\zeta_k^{-1} d\zeta_k = 2 \pi i, \ee where $\partial D_k$ denotes the boundary of
each electron disk. Due to this fact, the computation of the flux (\ref{flux})
must take into account all different configurations for $\vec A(z)$ \cite{fqhe2}.
Assuming that all three configurations enter with the same weight, and using
unities in which $q=1$ (so that the flux quanta becomes $\Phi_0=1$), the total
flux turns out to be \beq \label{fluxto} \Phi =\oint_{\partial D_1\cup\partial
D_2}A(z)dz+\oint_{\partial D_1\cup\partial D_2}\lp A(z)dz-\frac{i}{4\pi}\zeta_1^{-
1}d\zeta_1-
\frac{i}{4\pi}\zeta_2^{- 1}d\zeta_2\rp \nonumber \\
+\oint_{\partial D_1\cup\partial D_2}\lp A(z)dz-\frac{i}{4\pi}\zeta_1^{-
1}d\zeta_1\rp + \oint_{\partial D_1\cup\partial D_2}\lp A(z)dz-
\frac{i}{4\pi}\zeta_2^{-1}d\zeta_2\rp. \eeq Using then the flux quantization
condition \be \oint_{\partial D_1\cup\partial D_2}A(z) \, dz=n; \quad n=1, 2,
\dots , \ee we get \be \Phi = 4n - 2. \ee If we compute now the relation $\nu =$
{\it number of electrons/number of flux quanta}, we get \be \label{filling}
\nu=\frac{2}{4n-2}=\frac{1}{2n-1}. \ee This experimental setup is used in the
study of the Fractional Quantum Hall Effect (FQHE) \cite{fqhe3}, and in this
context the quantity $\nu$ is called the {\em filling factor}. The above result
coincides with the experimental one if we consider that the interactions between
electrons on the confining plane are pair-dominated \cite{fqhe4}.

\section{Conclusions}

We have shown in this paper that, by using the extended spacetime interpretation
of Hawking and Ellis together with Wheeler's idea of charge without charge, the KN
solution exhibits properties that are quite similar to those presented by an
electron, paving the way for the construction of an electron model \cite{bur}. The
first important point is that, due to its topological structure, the extended KN
solution admits the presence of spacetime spinorial structures. As a consequence,
the KN solution can naturally be represented in terms of spinor variables of the
Lorentz group SL(2, $\mathbf{C}$). The evolution of the KN state vector so
obtained is then shown to be governed by the Dirac equation. Another important
point is that this model provides a topological explanation for the concepts of
mass, charge and spin. Mass can be interpreted as made up of gravitation, as well
as rotational and electromagnetic energies, since all of them enter its
definition. Charge, on the other hand, is interpreted as arising from the multi-
connectedness of the spatial section of the KN solution. In fact, according to
Wheeler, from the point of view of an asymptotic observer, a trapped electric
field is indistinguishable from the presence of a charge distribution in
spacetime. Finally, spin can be consistently interpreted as an internal rotational
motion of the infinitesimally sized $U_2$ surface. Besides these properties, we
have also shown that this model can provide  explanations to not well-understood
phenomena of solid state physics, as for example the fractional quantum Hall
effect. It is important to remark once more that the topology ensued by the
Hawking and Ellis interpretation of the Kerr-Newman solution leads actually to
fundamental states with spin 1/2 only. Observe, for example, that the Kerr-Newman
solution presents four independent states, a typical property of the electron-
positron system. Notice, however, that it is possible to construct states with
higher spins by considering composed states, with the spin 1/2 solution as
building blocks. This, of course, changes the topology of the whole solution.
Higher-spin states are therefore possible, but then the manifold $\mathcal{M}$
will exhibit a different topology from the presented here. Finally, we would like
to remark that the metric does not fix the sign of $r$ at each side of the $r=0$
surface. It is actually a matter of convention, an arbitrary choice of the
observer.

\section*{Acknowledgments}
The authors would like to thank R.\ Aldrovandi, D.\ Galetti and Yu.\ N.\ Obukhov
for useful discussions. They would like to thank also FAPESP-Brazil, CNPq-Brazil,
COLCIENCIAS-Colombia and CAPES-Brazil for financial support.

\begin{appendix}
\section*{Appendix A. Lorentz Group and Parity Transformations}
\label{a}

As is well known, the complexified Lie algebra of the Lorentz group SL(2,
$\mathbf{C}$) is isomorphic to the complexified Lie algebra of the group SU(2)
$\otimes$ SU(2). In fact, denoting by $J_i$ and $K_i$ respectively the generators
of infinitesimal rotations and boost transformations in {\bf M}, with $i, j, k =
x, y, z$, the complex generators \be \label{cuatro} A_i=\frac{1}{2}\lp J_i +
iK_i\rp \quad {\rm and} \quad A^\dagger_i=\frac{1}{2}\lp J_i - iK_i\rp \ee are
known to satisfy, each one, the SU(2) Lie algebra \cite{ramond}: \beq
\lb A_i, A_j\rb &=&i \, \epsilon_{ijk} \, A_k \nonumber \\
\lb A^\dagger_i, A^\dagger_j\rb &=&i \, \epsilon_{ijk} \, A^\dagger_k. \nonumber
\eeq Furthermore, they satisfy also \be \lb A_i, A^\dagger_j\rb = 0, \ee which
shows that they are independent, or equivalently, a direct product. The two
Casimir operators $A_i A_i$ and $A_i^\dagger A_i^\dagger$ are also known to
present respectively eigenvalues $n(n+1)$ and $m(m+1)$, with $n, m = 0, 1/2, 1,
3/2, \dots$. Thus, each representation can be labeled by the pair $(n, m)$. Now,
as a simple inspection shows, under a parity transformation, \be J_i \rightarrow
J_i \quad {\rm and} \quad \quad K_i \rightarrow - K_i. \ee Therefore, the
generators $A_i$ and $A_i^\dagger$ are easily seen to be related by a parity
transformation \cite{ramond}.

On the other hand, it is clear from the KN metric (\ref{metric}) in {\bf M}, which
is written in terms of the coordinates $(t,r,\theta,\phi)$, that the KN solution
in {\bf M'} is written in terms of $(t,-r,\theta,\phi)$. Then, the gradient
function $\nabla r$ changes sign in {\bf M'}, making its Cartesian coordinate
system, with origin in the center of the disk, to present negative unitary
vectors. This is so because the unitary Cartesian vectors are perpendicular to the
$r$ = {\em constant} surfaces. The two sides of the KN solution, therefore, are
related by a parity transformation. The conclusion is that the relationship
between {\bf M} and {\bf M'} is the same as that between $A_i$ and $A_i^\dagger$.
This justifies the use in {\bf M} of Weyl spinors transforming under the $(1/2,0)$
representation, and the use in {\bf M'} of Weyl spinors transforming under the
$(0,1/2)$ representation of the Lorentz group.

\section*{Appendix B. Topological Properties of $\mathbf{K}^3$ }
\label{b}

Let $F: {\mathcal M}^3\cup\{\pm\infty\} \longrightarrow
\mathbf{R}^3\cup\{\infty\}$ be a function from the metric space ${\mathcal M}^3$
(plus two points at infinity) to a 3-dimensional Euclidean space (plus one point
at infinity). The function is defined by: \beq\label{eqB1}
F(r,\theta,\phi)&=&(e^r,\theta,\phi) \nonumber\\
F(-\infty,\theta,\phi)&=&(0,\theta,\phi) \nonumber\\
F(\infty,\theta,\phi)&=&(\infty,\theta,\phi) \nonumber. \eeq The space
$\mathbf{R}^3\cup\{\infty\}$ is the Alexandroff's one-point compactification of
$\mathbf{R}^3$, which is topologically equivalent to $S^3$. The function $F$ takes
the  surface $r=0$ of the KN solution into a sphere of unit radius, centered at
$(0,0,0)$. The singular ring is mapped into the equator of the sphere. The
function $F$ is continuous at any point  $c\in {\mathcal M}^3\cup\{\pm\infty\}$ if
for any positive real number  $\varepsilon$, there exists a positive real number
$\delta$ such that for all  $p\in{\mathcal M}^3\cup\{\pm\infty\}$ satisfying
$d_\gamma(p,c)<\delta$, the  inequality $d_\eta(F(p),F(c))<\varepsilon$ is also
satisfied. As the distance function defined by the metric $\gamma_{ij}$ of
${\mathcal M}^3$, given by (\ref{3metric}), is  finite everywhere, the continuity
condition is valid for every $p\in{\mathcal  M}^3$. However, as the border of the
disk $r=0$ is a single point of ${\mathcal M}^3\cup\{\pm\infty\}$, the function is
not one-to-one. In fact, the point ($r=0$, $\theta=\pi/2$) is mapped into the
equator of the unit sphere. Now,  since $S^3$ is compact, and $F^{-1}$ is
continuous, ${\mathcal M}^3\cup\{\pm\infty\}$ is also compact. If we excise an
{\it open} set from $S^3$ in the form of a solid torus (without boundary) around
the equator of the unitary sphere, we are left with a closed subset of $S^3$. This
closed subset is also compact and has as boundary a 2-dimensional torus $T^2$.
Denoting by $T^3$ the solid torus and by $T_F^3$ its image under $F^{-1}$, the
function
\[
(F^{-1})':\mathbf{R}^3\cup\{\infty\}-T^3 \longrightarrow  {\mathcal
M}^3\cup\{\pm\infty\}-T_F^3
\]
is found to be continuous. As a consequence, ${\mathcal M}^3\cup\{\pm\infty\}-
T_F^3$ will also be compact. Now, let $(\alpha,\beta)$ be the coordinates of
$T^2=S^1\times S^1$, and consider the map $A: T^2 \longrightarrow T^2$ defined by
$A(\alpha,\beta)=(\alpha+\pi,\beta+\pi)$. The surface obtained by making $A(p)
\sim p$, and then by taking the quotient space $T^2/A$, is the Klein bottle $U_2$.
In  this way, every point $p$ of $U_2$ has a neighborhood in $S^3$ which is
homeomorphic to a Euclidean open ball. This implies that
$\{\mathbf{R}^3\cup\{\infty\}-T^3\}/A$ is a compact manifold without boundary.
Therefore,
\[
{\mathcal F}: \{\mathbf{R}^3\cup\{\infty\}-T^3\}/A\longrightarrow \{{\mathcal
M}^3\cup\{\pm\infty\}-T_F^3\}/\{(F^{-1})'\circ(A)\}=\mathbf{K}^3
\]
will be a continuous function from a compact space, which means essentially that
$\mathbf{K}^3$ is also a compact space. Furthermore, since $\mathcal F$ is
continuous, it maps every open set of $\{\mathbf{R}^3\cup\{\infty\}-T^3\}/A$ into
an open set of $\mathbf{K}^3$, which implies that $\mathbf{K}^3$ has no boundary:
$\partial\mathbf{K}^3=0$.
\end{appendix}



\begin{thebibliography}{31}

\bibitem{kerr}
Kerr, R. P. (1963). {\it Phys. Rev. Lett.} {\bf 11}, 237.

\bibitem{newman}
Newman, E. T., and Janis, A. I. (1965). {\it J. Math. Phys.} {\bf 6}, 915.

\bibitem{newman2}
Newman, E. T. {\it et al} (1965). {\it J. Math. Phys.} {\bf 6}, 918.

\bibitem{hellis}
Hawking, S. W., and Ellis, G. F. R. (1973). {\it The Large Scale Structure of
Space-Time} (Cambridge University Press, Cambridge) p. 161.

\bibitem{lopez}
Lopez, C. A. (1984). {\it Phys. Rev.} {\bf D30}, 313; Lopez, C. A. (1992). {\it
Gen. Rel. Grav.} {\bf 24}, 285.

\bibitem{otro}
Israelit, M., and Rosen, N. (1995). {\it Gen. Rel. Grav.} {\bf 27}, 153.

\bibitem{israel}
Israel, W. (1970). {\it Phys. Rev.} {\bf D2}, 641.

\bibitem{barut1}
Barut, A. O., and Bracken, A. J. (1981). {\it Phys. Rev.} {\bf D23}, 2454.

\bibitem{barut2}
Barut, A. O., and Thacker, W. (1985). {\it Phys. Rev.} {\bf D31}, 1386.

\bibitem{mtw}
Wheeler, J. A. (1962). {\it Geometrodynamics} (Academic Press, New York).

\bibitem{fs}
Friedman, J. L., and Sorkin, R. (1980). {\it Phys. Rev. Lett.} {\bf 44}, 1100.

\bibitem{wald}
Wald, R. M. (1984). {\it General Relativity} (The University of Chicago Press,
Chicago) p. 289.

\bibitem{komar}
Komar, A. (1959). {\it Phys. Rev.} {\bf 113}, 934.

\bibitem{ohanian}
Ohanian, H., and Ruffini, R. (1994). {\it Gravitation and Spacetime} (Norton \&
Company, New York) p. 396.

\bibitem{landau}
Landau, L. D., and Lifshitz, E. M. (1975). {\it The Classical Theory of Fields}
(Pergamon, Oxford).

\bibitem{carter}
Carter, B. (1968). {\it Phys. Rev.} {\bf 174}, 1559.

\bibitem{carmo}
do Carmo, M. P. (1976). {\it Differential Geometry of Curves and Surfaces}
(Prentice-Hall, New Jersey).

\bibitem{punsly}
Punsly, B. (1987). {\it J. Math. Phys.} {\bf 28}, 859.

\bibitem{hen}
Hendriks, H. (1977). {\it Bull. Soc. Math. France Memoire} {\bf 53}, 81; see \S
4.3

\bibitem{st}
Steenrod, N. (1974). {\it The Topology of Fibre Bundles} (Princeton University
Press, New Jersey) p. 134.

\bibitem{pr}
Penrose, R., and Rindler, W. (1984). {\it Spinors and Spacetime}, Vol. 1
(Cambridge University Press, Cambridge) p. 43.

\bibitem{gross}
Gross, F. (1993). {\it Relativistic Quantum Mechanics and Field Theory} (Wiley,
New York) p. 147.

\bibitem{edm}
Commins, E. D. {\it et al} (1994). {\it Phys. Rev.} {\bf A50}, 2960.

\bibitem{buri2}
Burinskii, A., and Magli, G. (2000). {\it Phys. Rev.} {\bf D61}, 044017.

\bibitem{norma}
Luosto, C. O., and Sanchez, N. (1992). {\it Nucl. Phys.} {\bf B383}, 377.

\bibitem{fqhe1}
Petry, H. R. (1979). {\it J. Math. Phys.} {\bf 20}, 231.

\bibitem{fqhe2}
Avis, S. J., and Isham, C. J. (1979). {\it Nucl. Phys.} {\bf B156}, 441.

\bibitem{fqhe3}
Verlinde, E. (1992). {\it Quantum Hall Effect}, ed. by M. Stone (World Scientific,
Singapore).

\bibitem{fqhe4}
Asselmeyer, T., and Keiper, R. (1995). {\it Ann. Phys.} (Lpz) {\bf 4}, 739;
Asselmeyer, T., and Hess, G. (1995). {\it Fractional Quantum Hall Effect,
Composite Fermions and Exotic Spinors} (cond-mat/9508053).

\bibitem{bur}
Burinskii, A. (2003). {\it Phys. Rev.} {\bf D68}, 105004.

\bibitem{ramond}
Ramond, P. (1989). {\em Field Theory: A Modern Primer}, 2nd edition (Addison-
Wesley, Redwood).
\end{thebibliography}
\end{document}